\documentclass[8.5pt,twoside,twocolumn]{article}
\usepackage[super,sort&compress,comma]{natbib} 
\usepackage[version=3]{mhchem}
\usepackage{balance}
\usepackage{times,mathptm}
\usepackage{sectsty}
\usepackage{graphicx} 
\usepackage{lastpage}
\usepackage[format=plain,justification=raggedright,singlelinecheck=false,font=small,labelfont=bf,labelsep=space]{caption} 
\usepackage{fancyhdr}
\begin{document}

\thispagestyle{plain}
\fancypagestyle{plain}{
\renewcommand{\headrulewidth}{1pt}}
\renewcommand{\thefootnote}{\fnsymbol{footnote}}
\renewcommand\footnoterule{\vspace*{1pt}%
\hrule width 3.4in height 0.4pt \vspace*{5pt}} 
\setcounter{secnumdepth}{5}

\makeatletter 
\renewcommand\@biblabel[1]{#1}            
\renewcommand\@makefntext[1]%
{\noindent\makebox[0pt][r]{\@thefnmark\,}#1}
\makeatother 
\renewcommand{\figurename}{\small{Fig.}~}
\sectionfont{\large}
\subsectionfont{\normalsize} 
\fancyhead{}
\setlength\bibsep{1pt}

\twocolumn[
  \begin{@twocolumnfalse}
\noindent\LARGE{\textbf{Micro helical polymeric structures produced by variable voltage direct electrospinning}}
\vspace{0.6cm}

\noindent\large{\textbf{S. P. Shariatpanahi\textit{$^{a}$}, A. Iraji zad$^{\ast}$\textit{$^{a,b\ddag}$},I. Abdollahzadeh\textit{$^{a}$},R. Shirsavar\textit{$^{a}$},D. Bonn\textit{$^{c,d}$} and R. Ejtehadi \textit{$^{a}$}}}\vspace{0.5cm}


\noindent \textbf{\small{DOI:10.1039/C1SM06009K}}
 \end{@twocolumnfalse} \vspace{0.6cm}

  ]

\noindent\textbf{ Direct near field electrospinning is used to produce very long helical
polystyrene microfibers in water. The pitch length of helices can
be controlled by changing the applied voltage, allowing to produce
both micro springs and microchannels. Using a novel high frequency variable voltage electrospinning method we found the helix formation speed and compared the experimental buckling frequency to theoretical expressions for viscous and elastic
buckling. Finally we showed that the new method can be used to produce new  periodic micro and nano structures.
}
\section*{}
\vspace{-1cm}

\footnotetext{\textit{$^{a}$~Department of Physics, Sharif University of Technology, P.O. Box 11155-9161, Tehran, Iran. }}
\footnotetext{\textit{$^{b}$~Institute for Nanoscience and Nanotechnology (INST), Sharif University of Technology, P.O. Box 11155-9161, Tehran, Iran.}}
\footnotetext{\textit{$^{c}$~Laboratoire de Physique Statistique, CNRS UMR 8550, Ecole Normale Superieure, 24 Rue Lhomond,
75231 Paris Cedex 05, France.}}
\footnotetext{\textit{$^{d}$~Van der Waals-Zeeman Institute, University of Amsterdam,
Valckenierstraat 65, 1018 XE Amsterdam, The Netherlands.}}


\footnotetext{\ddag~Fax: 21 6602 2711; Tel: 21 6616 4513; E-mail: irajizad@sharif.edu}
Buckling of liquid or solid ropes, in which the falling rope on
the surface forms a helical shape, has been investigated both
theoretically and experimentally on stationary \cite{Daniel1,Daniel2,Daniel3} and moving \cite{discrete,Morris,lateral,lateral1}surfaces for typical length scales of
centimetres to meters . Recently it was
shown that using near field electrospinning
\cite{reneker_buckling,pottery,direct_writing,EPJ} the same buckling phenomenon can be used
at very small length scales to produce micro helical structures.
In these experiments an electrically charged liquid jet is
accelerated using a high voltage potential directly along its axis
toward a collector surface. Before the jet experiences any
electrical instability\cite{reneker_instability,yarin} (as happens in
usual electrospinning), it buckles and solidifies on the
collector surface.

These fine structures can find many applications in technological
fields like microelectronics, MEMS(Micro Electro Mechanical
Systems), Microfluidics etc., but so far no detailed control over
the fine structures has been possible. Kim et. al. used the
method to directly electrospin polyethylene oxide (PEO) solutions;
as the jet solidifies and buckles over a sharp electrode tip, hollow
coiled cylindrical structures with lengths of a few micrometers
were produced \cite{pottery}. Han et. al. showed that with the
same process fine polystyrene helical fibers can be produced by
using water as collector surface; the latter speeds up the
solidification of the jet \cite{reneker_buckling}. Using the same method, here we show how the buckling process can be
controlled so that a wide range of morphologies can be obtained:
from straight fibers to helices and even micro channels.
Furthermore for the first time we use a high frequency variable
high voltage source that allows to control and measure the
buckling frequency and the jet velocity. These measurements allow
 to compare the experimental results with existing theories for buckling.

\begin{figure}
\begin{center}
\includegraphics[scale=0.35]{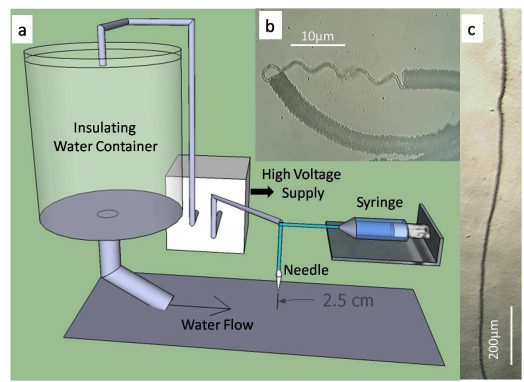}
\caption{(a) Schematic of the direct electrospinning experiment (b,c)
long hollow coiled micro channels produced by direct electrospinning using water flow as collector 
\label{fig:setup}}
\end{center}
\end{figure}

Fig.\ref{fig:setup}(a) shows the set up of the experiment; a
high voltage potential difference is applied between the needle of
a syringe and a bath of gently flowing water, $2.5 cm$ apart. The
straight jet of polymer solution coming out from a syringe needle
is collected over the water surface and moves with the flow. The
constant flow is made using a water container (the cylindrical
container in the figure) with the high voltage electrode kept
in the water close to the surface; the height of the water and consequently the
flow velocity do not change considerably over the experiment time
window.

As the polymer solution we used a Poly Styrene (PS) (Mw=350,000
a.m.u.) in Dimethyl Formamide(DMF); both from Sigma Aldrich. The solution solidifies
rapidly as it touches the water so that after the process we
obtain fine helical PS fibers. The process was controlled so that
a long (up to tens of centimetres) buckled helical fiber is
obtained (Fig.\ref{fig:setup}(b,c)). 

\begin{figure}[t]
\begin{center}
\includegraphics[scale=0.35]{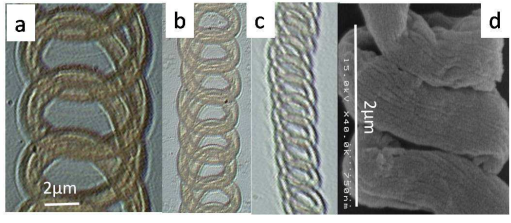}
\caption{Buckled helical structures: Optical microscope images with the same magnifications for: (a) $C=25 wt \%,V=4kV$ (b) $C=15  wt \%,V=4kV$ (c) $C=10 wt \% ,V=4kV$ and SEM image for (d) $C=10 wt \%, V=6kV$ \label{fig:ds}}
\end{center}
\end{figure}

Collecting the fiber on
a glass slide we could subsequently visualize it using optical
microscopy. Figures \ref{fig:ds} (a,b and c) show the helices produced for different polymer concentrations $C$ in constant applied  voltage $V=4kV$.  The most important observation that the fiber size can be controlled by changing the
polymer concentration (table \ref{tab:diameter}); the range we obtain here is from a few
micrometers to a few hundreds of nanometers. 

\begin{table}
\begin{center}
\caption{Different fiber diameters for different solution concentrations $C$ (Fig \ref{fig:ds}) \label{tab:diameter}}
\begin{tabular}{c c c}
\hline
 $10 wt \%$ & $15 wt \%$ & $25 wt \%$\\ \hline
 $0.7 \pm 0.1 \mu m$ & $1.1 \pm 0.2 \mu m$ & $1.8 \pm 0.2 \mu m$\\ \hline
\end{tabular}
\end{center}
\end{table}

\begin{figure}
\begin{center}
\includegraphics[scale=0.3]{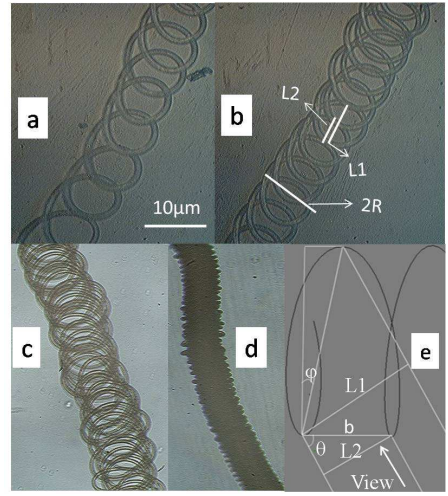}
\caption{Optical microscope images with the same magnifications of buckled helical fibers for: (a) $V=3
kV,C=25 wt \%$ (b) $V=4kV,C=25 wt \%$ (c)$V=5kV,C=25 wt \%$ (d)$V=6kV,C=25 wt \%$ and (e) The helix geometry\label{fig:bs}}
\end{center}
\end{figure}

Fig.\ref{fig:bs} shows the helices for different applied voltages $V$ in constant concentration $C=25 wt \%$. We find
that the pitch length of the helices can be controlled with
voltage from wide helices to channels formed by a dense helix. In addition we see that the fiber diameter remains almost constant as the voltage is varied.  The pitch length can be measured for $3D$ structured helices taking into account the direction of observation.  Fig.\ref{fig:bs}(e) shows a schematic of a helix
from its side view. An observer looking at an angle $\theta$ with
the axis of the helix measures the radius of the helix $R$, the
lengths $L_1$ and $L_2$. The pitch length $b$ of the helix is then
easily found from:
\begin{eqnarray}
&&L_1=\cos(\theta - \phi) \sqrt{(\frac{b}{2})^2+(2R)^2}\nonumber \\
&&L_2=\sin(\theta) b\nonumber\\
&&\tan(\phi)=\frac{b}{4R}
\label{eqn:obtain-b}
\end{eqnarray} 
Table \ref{tab:ds} shows the pitch lengths found for the different voltages.

\begin{table}
\begin{center}
\caption{Different helix pitch lengths for different voltages $V$ (Fig \ref{fig:bs}) \label{tab:ds}}
\begin{tabular}{c c c c}
\hline
 $3 kV$ & $4 kV$ & $5 kV$ & $6 kV$ \\ \hline
 $10 \pm 1 \mu m$ & $5 \pm 0.5 \mu m$ & $3.5 \pm 0.5 \mu m$ & $1.5 \pm 0.5 \mu m$\\ \hline
\end{tabular}
\end{center}
\end{table}

\begin{figure}
\begin{center}
\includegraphics[scale=0.35]{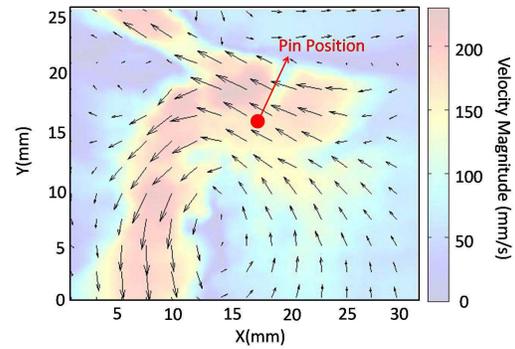}
\end{center}
\caption{Flow velocity field caused by ion wind over the water
surface 
\label{fig:velocity}}
\end{figure}

\begin{figure}[t]
\includegraphics[scale=0.35]{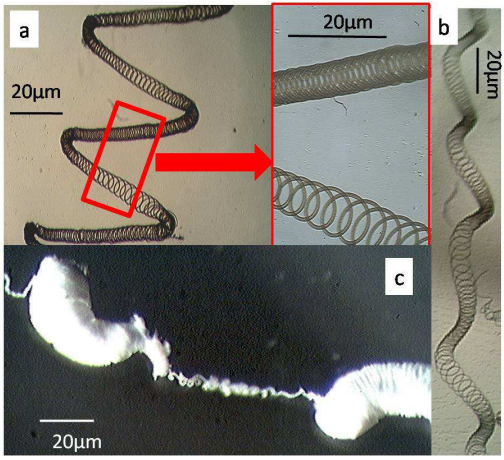}
\caption{Helical structure produced by applying a 3kV DC high
voltage added to a sinusoidal 1.5 kV p-p variable voltage with
frequencies (a) 200 Hz (b) 1 kHz using a $C= 25 wt \%$ solution. (c) New periodic structure produced by direct
electrospinning using a square wave $5kV$ high voltage with frequency $100Hz$ 
.
\label{fig:variable}}
\end{figure}
Since we are clearly out of the range of electrical
electrospinning instabilities \cite{reneker_instability}, it appears that the
coiling we observe is purely mechanical. Recently, coiling
instabilities have been studied in detail for both liquids and
solids, and for both cases the coiling frequencies are well
understood by now \cite{AnnRev}.  To be able to confront our
experiments to the theoretical predictions of Habibi et. al.
\cite{Daniel1,Daniel2} we should determine the buckling frequency. In
principle, this frequency can be deduced from the length scale of
the coils and knowing the lateral velocity of advection of the
buckled fiber\cite{reneker_buckling}; however in our case the ion
wind produced by the applied high voltage in air constitutes a problem \cite{ion_wind}. In the experiments, we
observed a surface motion caused by the ion wind by recording the motion of added tracer particles smaller than
$100 \mu m $ with a high speed camera (Casio EX-F1). Using Particle Image Velocimetry (PIV) we subsequently measure their
velocity at the surface of the water. As Fig.\ref{fig:velocity}
shows it is not possible here to measure the advection speed
accurately, because the ion wind causes a complicated velocity
field on the water surface.  In
addition the velocity of the flow is not necessarily the same as
the velocity of the buckled fiber as it remains attached to the
jet that falls from the orifice.

To be able to find the buckling frequency we used a novel method
using a sinusoidal variation of the high voltage source superposed
onto the DC voltage. As shown in Fig \ref{fig:variable} the
resulted buckled helices have variable pitch lengths and radius
with the helix axis bended periodically in specific points. The
variation of the pitch length and radius is due to the variable
applied voltage during the fiber formation; while the bending of the helix, observed to occur after the
buckling process, is caused by the variation of the spring
constant along the helix axis. Since the bending points have the
same axis length period as the pitch length variations we can
consider these as the points with the same pitch length. In
addition we observed the ion wind of the high frequency
($>100Hz$) voltage did not cause any flow on the water surface, so that the velocity of the water flow is constant while a considerable number of loops are formed. Since the velocity is constant, the axis length of the produced helix is proportional to time and
we can safely assume that the length along the helix axis in each
period is proportional to time the loops formed. So each photo
works as a fast camera recording the buckling process and the
buckling frequency is easily found by measurement of the pitch
length. This allows to compare with the publications of Habibi et. al. \cite{Daniel1,Daniel2,Daniel3}

\begin{figure}
\begin{center}
\includegraphics[scale=0.23]{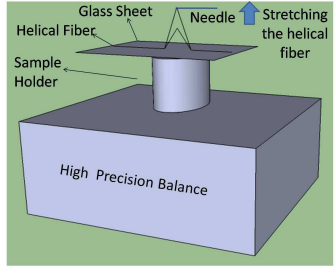} 
\end{center}
\caption{Schematic of the experiment for measuring the spring constant of the helical fibers using a needle for stretching the helical structure over a precision balance\label{fig:module-measure}}
\end{figure}

\begin{figure}
\includegraphics[scale=0.4]{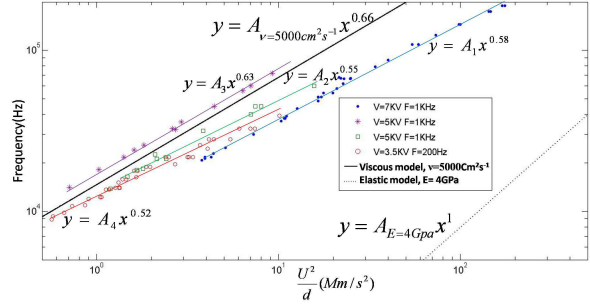}
\caption{Comparison between the theoretical models
\cite{Daniel1,Daniel2} and the experimental results for different
DC voltages $V$ and different variable $1.5kV P-P$ voltage
frequencies $F$ \label{fig:result}}
\end{figure}

To compare the experiments with the theory for buckling we
introduce the following electrical $F_e$, inertial $F_I$, viscous
$F_\nu$ and elastic $F_E$ forces per unit length to investigate
the dynamics \cite{pottery,Daniel1,Daniel2}. Then:
\begin{eqnarray}
&&F_e\sim \epsilon_0 e^2 r\nonumber\\
&& F_I\sim\frac{\rho r^2 U^2}{R}\nonumber\\
&& F_{\nu}\sim\frac{ \rho \nu r^4 U}{R^4}\nonumber\\
&& F_E\sim \frac{E r^4}{R^3}
\end{eqnarray}
Where $r$ and $U$ are the radius and the velocity of the jet
before buckling, $\rho$, $\nu$ and $E$ are the density, kinematic
viscosity and the Young modulus of the jet, $R$ is the radius of
the buckled helix, $\epsilon_0$ is the permeability of vacuum and
$e$ is the electrical field. In our case the observed experimental
parameters are of $U\sim 1m/s,r\sim 1 \mu m, \rho\sim
10^3 Kg/m^3$, $e\sim 10^3 V/cm$ and $R\sim 10 \mu m$ for which
$F_e/ F_I\sim10^{-3}$. Habibi et. al. previously showed that for
these conditions, the inertial forces are dominant: we are in the
inertial regime of buckling. This result is different from previous work by Kim et. al. \cite{pottery} in which the electrical forces in the focused field near the sharp electrode tip are much larger than the inertial forces. The question remains then whether the
jet should be considered as a solid rope, or as a liquid filament.
For these two different cases, the buckling frequency for the
viscous ($\Omega_{I\nu}$) and elastic jets ($\Omega_{IE}$) are
given by \cite{Daniel1,Daniel2}.:
\begin{eqnarray}
\Omega_{I\nu}=\frac{1}{2 \pi}(\frac{(\pi U)^4}{\nu r^2})^{1/3}\nonumber\\
 \Omega_{IE}=\frac{1}{2 \pi} U^2(\frac{\rho}{d^2 E})^{1/2} \label{eqn:freq}
\end{eqnarray}
In which $d=2r$ is the diameter of the fiber. Although these
results are for buckling on a surface that does not move, it can
be shown that the lateral motion of the surface does not affect
the results significantly \cite{lateral}. Since during the
experiment the viscous PS solution changes to a solid elastic
material rapidly when it touches the water, to find out what the
proper model for the buckling is we have to measure both the
viscosity of the jet before the buckling and the Young's modulus
after this happens. The kinematic viscosity of the solution was
determined using a Physica MCR301 rheometer to be $5000 cm^{2}
s^{-1}$. For the Young's modulus measurement we found the spring
constant of the buckled helix using a needle for stretching the helical structure over a precision balance(Fig \ref{fig:module-measure}).
 The helical fibers showed elastic
behaviour for strains smaller than $25\%$ which was sufficient for
an accurate measurement of the spring constant. Measuring the
spring pitch length, helix radius and fiber diameter, we find the
Young's modulus to be almost $E=4 GPa$, very similar to the usual
bulk PS material ($3.5 GPa$).
These measurements allow us to confront the experimental results to
the predictions for viscous and elastic coiling.
Fig\ref{fig:result} shows the buckling frequency vs $U^2/d$ on a
log-log scale. The elastic model predicts a slope of unity,
whereas for viscous coiling the slope would be $2/3$ (see
Eqn. \ref{eqn:freq}). For the experimental data, the slope of
 we find is always smaller than but comparable to $2/3$ which
 suggests that the buckling is more viscous than elastic in
 character. More quantitatively,
Fig \ref{fig:result} shows that the viscous model somewhat
overestimates most of the experimentally measured frequencies.
This can in fact be explained by assuming that the viscosity
during the buckling is a little higher than the bulk one
measured in the rheometer. This difference may be due to the thin ($\sim 1 \mu m$ diameter) jet
starting to solidify because of solvent evaporation. The importance of evaporation in these type of experiments is also reported by other groups \cite{pottery,direct_writing}.

As a bonus, the method employed here also allows to manufacture
different types of periodic structures. As an example Fig
\ref{fig:variable} (c) shows such a structure made by
applying square wave high voltage with $100Hz$ frequency and $5kV$
p-p amplitude. The structure consists of many hollow cavities connected periodically to each other. The cavities are produced by buckling and parallel fibers can be easily observable on their body. Since the parameter range in these experiments is rather large, one may
find a large number of novel structures; however this is beyond
the scope of the current paper.

In conclusion, we used a direct electrospinning
method to produce buckled helical PS fibers with controlled pitch
lengths. We used a high variable sinusoidal voltage added to the
usual electrospinning high DC voltage to measure the experimental
buckling frequency which showed a better agreement with the
theoretical viscous rope buckling model than with the elastic one.
Finally we introduced the new electrospinning technique as a
method to produced new fine periodic micro structures whose
characteristics can be controlled by experimental parameters such
as the polymer concentration or the applied voltage.

\footnotesize{
\providecommand{\noopsort}[1]{}\providecommand{\singleletter}[1]{#1}%
\providecommand*{\mcitethebibliography}{\thebibliography}
\csname @ifundefined\endcsname{endmcitethebibliography}
{\let\endmcitethebibliography\endthebibliography}{}

 }


\begin{mcitethebibliography}{15}
\providecommand*{\natexlab}[1]{#1}
\providecommand*{\mciteSetBstSublistMode}[1]{}
\providecommand*{\mciteSetBstMaxWidthForm}[2]{}
\providecommand*{\mciteBstWouldAddEndPuncttrue}
  {\def\EndOfBibitem{\unskip.}}
\providecommand*{\mciteBstWouldAddEndPunctfalse}
  {\let\EndOfBibitem\relax}
\providecommand*{\mciteSetBstMidEndSepPunct}[3]{}
\providecommand*{\mciteSetBstSublistLabelBeginEnd}[3]{}
\providecommand*{\EndOfBibitem}{}
\mciteSetBstSublistMode{f}
\mciteSetBstMaxWidthForm{subitem}
{(\emph{\alph{mcitesubitemcount}})}
\mciteSetBstSublistLabelBeginEnd{\mcitemaxwidthsubitemform\space}
{\relax}{\relax}

\bibitem[Habibi \emph{et~al.}(2010)Habibi, Rahmani, Bonn, and Ribe]{Daniel1}
M.~Habibi, Y.~Rahmani, D.~Bonn and N.~M. Ribe, \emph{Phys. Rev. Lett.}, 2010,
  \textbf{104}, 074301\relax
\mciteBstWouldAddEndPuncttrue
\mciteSetBstMidEndSepPunct{\mcitedefaultmidpunct}
{\mcitedefaultendpunct}{\mcitedefaultseppunct}\relax
\EndOfBibitem
\bibitem[Maleki \emph{et~al.}(2004)Maleki, M.Habibi, Golestanian, Ribe, and
  Bonn]{Daniel2}
M.~Maleki, M.Habibi, R.~Golestanian, N.~M. Ribe and D.~Bonn, \emph{Phys. Rev.
  Lett.}, 2004, \textbf{93}, 214502\relax
\mciteBstWouldAddEndPuncttrue
\mciteSetBstMidEndSepPunct{\mcitedefaultmidpunct}
{\mcitedefaultendpunct}{\mcitedefaultseppunct}\relax
\EndOfBibitem
\bibitem[Habibi \emph{et~al.}(2007)Habibi, Ribe, and Bonn]{Daniel3}
M.~Habibi, N.~M. Ribe and D.~Bonn, \emph{Phys. Rev. Lett.}, 2007, \textbf{99},
  154302\relax
\mciteBstWouldAddEndPuncttrue
\mciteSetBstMidEndSepPunct{\mcitedefaultmidpunct}
{\mcitedefaultendpunct}{\mcitedefaultseppunct}\relax
\EndOfBibitem
\bibitem[Bergou \emph{et~al.}(2010)Bergou, Audoly, Vouga, Wardetzky, and
  Grispun]{discrete}
M.~Bergou, B.~Audoly, E.~Vouga, M.~Wardetzky and E.~Grispun, \emph{ACM
  Transactions on Graphics}, 2010, \textbf{29},
  doi:10.1145/1778765.1778853\relax
\mciteBstWouldAddEndPuncttrue
\mciteSetBstMidEndSepPunct{\mcitedefaultmidpunct}
{\mcitedefaultendpunct}{\mcitedefaultseppunct}\relax
\EndOfBibitem
\bibitem[Morris \emph{et~al.}(2008)Morris, Dawes, Ribe, and Lister]{Morris}
W.~Morris, J.~H.~P. Dawes, N.~M. Ribe and J.~R. Lister, \emph{Phys. Rev. E},
  2008, \textbf{77}, 066218\relax
\mciteBstWouldAddEndPuncttrue
\mciteSetBstMidEndSepPunct{\mcitedefaultmidpunct}
{\mcitedefaultendpunct}{\mcitedefaultseppunct}\relax
\EndOfBibitem
\bibitem[Ribe \emph{et~al.}(2006)Ribe, Lister, and Chiu-Webster]{lateral}
N.~M. Ribe, J.~R. Lister and S.~Chiu-Webster, \emph{Phys. Fluids.}, 2006,
  \textbf{18}, 124105\relax
\mciteBstWouldAddEndPuncttrue
\mciteSetBstMidEndSepPunct{\mcitedefaultmidpunct}
{\mcitedefaultendpunct}{\mcitedefaultseppunct}\relax
\EndOfBibitem
\bibitem[Chiu-Webster and Lister(2006)]{lateral1}
S.~Chiu-Webster and J.~R. Lister, \emph{J. Fluid. Mech.}, 2006, \textbf{569},
  89\relax
\mciteBstWouldAddEndPuncttrue
\mciteSetBstMidEndSepPunct{\mcitedefaultmidpunct}
{\mcitedefaultendpunct}{\mcitedefaultseppunct}\relax
\EndOfBibitem
\bibitem[Han \emph{et~al.}(2007)Han, Reneker, and Yarin;]{reneker_buckling}
T.~Han, D.~H. Reneker and A.~L. Yarin;, \emph{Polymer}, 2007, \textbf{48},
  6064\relax
\mciteBstWouldAddEndPuncttrue
\mciteSetBstMidEndSepPunct{\mcitedefaultmidpunct}
{\mcitedefaultendpunct}{\mcitedefaultseppunct}\relax
\EndOfBibitem
\bibitem[Kim \emph{et~al.}(2010)Kim, Lee, Kim, and M.Mahadevan]{pottery}
H.~Y. Kim, M.~Lee, K.~J. P.~S. Kim and M.Mahadevan, \emph{Nano. Lett.}, 2010,
  \textbf{10}, 2138\relax
\mciteBstWouldAddEndPuncttrue
\mciteSetBstMidEndSepPunct{\mcitedefaultmidpunct}
{\mcitedefaultendpunct}{\mcitedefaultseppunct}\relax
\EndOfBibitem
\bibitem[Wang \emph{et~al.}(2010)Wang, Zheng, Li, Wang, and
  Sun]{direct_writing}
H.~Wang, G.~Zheng, W.~Li, X.~Wang and D.~Sun, \emph{Appl. Phys. A}, 2010,
  \textbf{102}, 457\relax
\mciteBstWouldAddEndPuncttrue
\mciteSetBstMidEndSepPunct{\mcitedefaultmidpunct}
{\mcitedefaultendpunct}{\mcitedefaultseppunct}\relax
\EndOfBibitem
\bibitem[Yu \emph{et~al.}(2008)Yu, Qiu, Zha, Yu, Yu, Rafique, and Yin]{EPJ}
J.~Yu, Y.~Qiu, X.~Zha, M.~Yu, J.~Yu, J.~Rafique and J.~Yin, \emph{European.
  Polymer. Journal}, 2008, \textbf{44}, 2838\relax
\mciteBstWouldAddEndPuncttrue
\mciteSetBstMidEndSepPunct{\mcitedefaultmidpunct}
{\mcitedefaultendpunct}{\mcitedefaultseppunct}\relax
\EndOfBibitem
\bibitem[Reneker \emph{et~al.}(2000)Reneker, Yarin, Fong, and
  Koombhongse]{reneker_instability}
D.~H. Reneker, A.~L. Yarin, H.~Fong and S.~Koombhongse, \emph{J. Appl. Phys.},
  2000, \textbf{87}, 4531\relax
\mciteBstWouldAddEndPuncttrue
\mciteSetBstMidEndSepPunct{\mcitedefaultmidpunct}
{\mcitedefaultendpunct}{\mcitedefaultseppunct}\relax
\EndOfBibitem
\bibitem[Yarin \emph{et~al.}(2001)Yarin, Koombhongse, and Reneker]{yarin}
A.~L. Yarin, S.~Koombhongse and D.~H. Reneker, \emph{J. Appl. Phys.}, 2001,
  \textbf{89}, 3018\relax
\mciteBstWouldAddEndPuncttrue
\mciteSetBstMidEndSepPunct{\mcitedefaultmidpunct}
{\mcitedefaultendpunct}{\mcitedefaultseppunct}\relax
\EndOfBibitem
\bibitem[Ribe \emph{et~al.}(to be published)Ribe, Habibi, and Bonn]{AnnRev}
N.~M. Ribe, M.~Habibi and D.~Bonn, \emph{Ann. Rev. Fluid.}, to be
  published\relax
\mciteBstWouldAddEndPuncttrue
\mciteSetBstMidEndSepPunct{\mcitedefaultmidpunct}
{\mcitedefaultendpunct}{\mcitedefaultseppunct}\relax
\EndOfBibitem
\bibitem[Kawamoto and Umezu(2005)]{ion_wind}
H.~Kawamoto and S.~Umezu, \emph{J. Phys. D: Appl. Phys.}, 2005, \textbf{38},
  887\relax
\mciteBstWouldAddEndPuncttrue
\mciteSetBstMidEndSepPunct{\mcitedefaultmidpunct}
{\mcitedefaultendpunct}{\mcitedefaultseppunct}\relax
\EndOfBibitem
\end{mcitethebibliography}
\end{document}